\documentclass[epjST]{svjour}
\usepackage{color}
\usepackage[english]{babel}
\usepackage[latin1]{inputenc}
\usepackage{graphicx,times}
\usepackage{amsmath,amssymb}
\usepackage{subfigure}
\usepackage{array}
\usepackage{graphics}
\usepackage{url}

\begin{document}
\title{Detecting and Describing Dynamic Equilibria in Adaptive Networks}
\author{Stefan Wieland \inst{1}\fnmsep\thanks{\email{swieland@cii.fc.ul.pt}} \and Andrea Parisi \inst{1} \and Ana Nunes \inst{1} }
\institute{Centro de F{\'\i}sica da Mat\'eria Condensada and
Departamento de F{\'\i}sica, Faculdade de Ci{\^e}ncias da Universidade de 
Lisboa, P-1649-003 Lisboa, Portugal
}
\abstract{
We review modeling attempts for the paradigmatic contact process (or \emph{SIS model}) on adaptive networks. Elaborating on one particular proposed mechanism of topology change (\emph{rewiring}) and its mean field analysis, we obtain a coarse-grained view of coevolving network topology in the stationary active phase of the system. Introducing an alternative framework applicable to a wide class of adaptive networks, active stationary states are detected, and an extended description of the resulting steady-state statistics is given for three different rewiring schemes. We find that slight modifications of the standard rewiring rule can result in either minuscule or drastic change of steady-state network topologies.
} 
\maketitle
\section{Introduction}
\label{s:intro}
In the past two decades, complex networks - multiple agents (\emph{nodes}) engaging in structured interactions (\emph{links} connecting nodes) - have become a prominent paradigm in tackling complex systems \cite{BA:2002,Dorogovtsev:2003,Newman:2003,Boccaletti:2006}. More recently, allowing for the coevolution of network structure and node-state dynamics has given rise to abundant literature on \emph{adaptive} networks \cite{Blasius:2008,Gross:2009}, capturing such diverse phenomena as the emergence of cooperation \cite{Ebel:2002,Zimmermann:2004,Zimmermann:2005,Szolnoki:2009,Fu:2009,Perc:2010}, opinion formation \cite{Holme:2006b,Kozma:2008,Vazquez:2008,Benczik:2009}, disease spreading \cite{Funk:2010,Zanette:2006,Gross:2006,Gusman:2009,Kiss:2012}, speciation \cite{Drossel:2001,Dieckmann:2004} and traffic flows \cite{Wang:2005,Lim:2007,Xie:2007}. While some contributions explore the respective phenomenology with individual-based simulations \cite{Zimmermann:2004,Szolnoki:2009,Herrera:2011,Zhang:2011}, others also focus on providing explanatory frameworks for observed dynamics \cite{Ebel:2002,Zimmermann:2005,Fu:2009,Vazquez:2008,Gross:2006,Gusman:2009,Lagorio:2011}.

Cyclic processes on adaptive networks, i.e. nodes going through a cyclic sequence of states while changing their connections, may yield adaptive networks displaying perpetual node-state and link dynamics \cite{Blasius:2008}. Those processes correspond to the active phase of the system, as opposed to the frozen phase where the system reaches a static equilibrium. As system dynamics often rely on the existence of \emph{active} links connecting nodes of different states, the frozen phase either manifests itself in a globally \emph{polarized} network (with all nodes carrying the same state) or a \emph{fragmented} network (featuring two or more connected components that are polarized, whereas
globally the network is not). Transitions to and between those absorbing states in adaptive networks have been dealt with using mean-field approximations or small perturbations of the fragmented state \cite{Vazquez:2008,Zanette:2006,Graeser:2009,Demirel:2011,Boehme:2011,Zschaler:2012}.

Node-state and link dynamics in the active phase of adaptive networks can be highly complex \cite{Bornholdt:1998,Holme:2006}, and stationary (as well as oscillatory) steady states have been shown to occur for a variety of frameworks both in Monte-Carlo (MC) simulations and in their approximate mean field descriptions \cite{Gross:2009}. For such networks in \emph{dynamic equilibrium} (DE), nodes undergo permanent state and degree evolution, while global statistics characterizing ensemble dynamics and network topology settle down to a steady state. Describing network statistics in DE is essential to understanding the relation between dynamics and network structure and to apply these ideas to real world examples.

Adaptive networks in the active phase have been treated analytically when timescales of node dynamics and topology evolution are separable \cite{Pacheco:2006,Segbroeck:2009,Segbroeck:2010}. For fast network dynamics in the context of evolutionary games, it was shown that the node states evolve according to an effective pay-off matrix that takes into account the equilibrium network properties \cite{Pacheco:2006,Segbroeck:2009}.

For the more general case that permits similar timescales, two equation-based frameworks taking the contact process as the underlying dynamics have so far been put forward. One is based on the pairwise formalism of \cite{Gross:2006} and related frameworks \cite{Gusman:2009,Kiss:2012,Zanette:2008}, where the time evolution of network motif densities is modeled up to the level of pairs, with description of highest-order motifs relying on the standard pair approximation moment closure assumption. This yields a low-dimensional set of ordinary differential equations (ODEs) that allows for analytic treatment, predicting system phases and pair densities in transients and DE with an accuracy limited by moment closure validity. Approximate phase diagrams were also derived in the scope of even simpler
effective mean-field descriptions \cite{Zanette:2008}. While this type of models gives a global account of the network in DE through averages that characterize steady-state dynamics, its low number of degrees of freedom precludes any detailed description of the network's topology. The approach used in \cite{Gross:2006} for the contact process was extended to a three-state model (the Susceptible-Infected-Recovered-Susceptible cycle) describing more realistic infection dynamics \cite{Shaw:2008}. Both models have stationary active phases, and simulations in DE converge to well-defined overall and state-specific degree distributions. In \cite{Shaw:2008}, we also find an attempt to describe the topology of the network by translating the pairwise dynamics of \cite{Gross:2006} to a degree-class formulation similarly to the ansatz in \cite{Vespignani:2001}. This approach yields a self-contained method to determine the degree distributions of all three node types, but fails to reproduce the observed output of MC simulations. Given however the infected degree distribution extracted from simulations, the two remaining distributions are accurately described.

Elaborating on a state- and degree-based compartmental formulation developed earlier \cite{Moreno:2004}, Marceau et al. in \cite{Marceau:2010} on the other hand modeled the time evolution of fractions of nodes with the same status and \emph{joint degree} (in a network with \(n\) possible node states, the joint degree \((k_1,k_2,...,k_n)\) of a node is the set of numbers \(k_j\) of its connections to nodes of type \(j \in \{1,2,...,n\}\)). In the spirit of the moment closure in the aforementioned ansatz, infection dynamics beyond a node's immediate neighborhood are approximated by a mean field that is computed en route and assumes no correlations beyond the level of next neighbors. A large set of coupled nonlinear ODEs ensues that defies analytic treatment. Instead, numerical integration yields the time evolution of the joint degree distribution, as well as of low-order network motif densities derived from it. This approach provides an alternative to stochastic simulations on networks for the description of the DE, with the same limitations due to transient's length and the additional loss of accuracy involved in the mean field approximation.

A third approach goes back to the original formulation of \cite{Gross:2006} and avoids
any moment closure approximation in the numerical integration of the ODEs 
by computing the triplet densities at each integration step through short bursts of MC simulations
on networks \cite{Gross:2008}. Much like \cite{Shaw:2008}, this hybrid approach gives a more precise ODE-based description of the system than its purely deterministic counterpart. It accurately reproduces the time evolution of the network's node and pair densities, and hence also the global phase diagram, but it brings no improvement regarding the analysis of the steady-state degree distributions.

Individual-based simulations of various adaptive networks in DE show that apart from global averages and degree-related probability distributions, a wide range of other topological measures settle down to an equilibrium. The most comprehensive description of DEs in frameworks like \cite{Marceau:2010,us:2011} is at the level of configuration model networks, implying that only the topology measures determined by the joint degree
distributions can be derived. To this point a more extensive account of steady-state topology has to resort to modeling the time evolution of the network's adjacency matrix or a related construct \cite{Guerra:2010}. In \cite{Bryden:2011} however, the steady-state community structure of a general class of adaptive networks in DE was characterized by a simple ODE for the network's modularity. 

In this paper, we will focus on a popular adaptive system: the Susceptible-Infected-Susceptible (SIS) model with rewiring by Gross et al. \cite{Gross:2006}. For that particular rewiring scenario, more predictions from their pairwise model are extracted. A novel analytic framework that stochastically models adaptive networks in DE is then introduced 
and applied to the contact process with rewiring, elaborating on results presented in \cite{us:2011}. Finally, different rewiring rules are considered, which in SIS terminology represent disease awareness under different conditions, and the proposed framework is used to show how the degree distributions in DE depend on them.

\section{SIS with Selective Rewiring in the Pairwise Model}
\label{s:sis}

As a simple cyclic process, the contact process on an adaptive network was proposed in \cite{Gross:2006} to model the spreading of a disease in a population without immunity, but with disease awareness. As in the traditional SIS model, infected nodes (\emph{I-nodes} of fraction \([I]\) of the total node number) transmit the disease to an adjacent susceptible node (an \emph{S-node} of the remaining node fraction of size \([S]\)) with rate \(p\), while recovering with rate \(r\). Additionally, S-nodes evade infection through \emph{selective rewiring} (SR) by cutting links to an infected neighbor with rate \(w\) and rewire it to a randomly selected S-node (double- and self-connections prohibited). 

This basic adaptive network with constant mean degree \(\langle k \rangle\) already displays a rich dynamical behavior in the active phase, featuring small regions of bistable and of stable oscillatory regimes right next to the dominating stable endemic phase \cite{Gross:2006}. All phases, as well as the time evolution of both node fractions and the various \emph{link densities} \([AB]\) (the number of \(AB\)-links per node connecting nodes of type \(A\) and \(B\), \(A,B \in \{S,I\}\)), are approximately captured by the pairwise framework proposed in \cite{Gross:2006}. While modeling the time evolution of the link densities requires tracking triplet densities \([ABC]\) (the number of triplets per node with the central node in stage \(B\) connected to a A- and C-node (\(A,B,C \in \{S,I\}\)), the latter are approximated by the standard pair moment closure assumption. Then, the evolution of the three degrees of freedom of the system in this description is given by 
\begin{align} \label{e:pm}
 \frac{d[I]}{dt} & =  p [SI] - r[I] \nonumber\\  
 \frac{d[II]}{dt} & =  p [SI]\left(\frac{[SI]}{[S]}+1\right) - 2r [II] \nonumber\\
 \frac{d[SS]}{dt} & =  [SI]\left( \left(w+r\right)- 2p\frac{[SS]}{[S]}\right),
\end{align}
with \(\langle k \rangle /2 =[SS]+[SI]+[II]\) \cite{Gross:2006}. 
Since
\[\langle k_S\rangle=\frac{2[SS]+[SI]}{[S]}\]
for the mean degree of the S-node subensemble (\emph{S-subensemble}) and
\[\langle k_I\rangle=\frac{2[II]+[SI]}{[I]}\]
for the I-node subensemble (\emph{I-subensemble}), Eqs. \ref{e:pm} also offer a coarse-grained view of coevolving 
network topology.
We find that in DE,
\[\langle k_S\rangle- \langle k_I\rangle=\frac{w}{p}-1,\]
regardless of imposed \(\langle k \rangle > 1\). Hence in steady state, the S-subensemble will have a larger mean degree than the I-nodes if and only if rewiring outperforms infection (see also \cite{Marceau:2010}). This is plausible, since the balance of mean degrees of both subensembles depends on whether on average a given SI-link is turned into a SS-link through rewiring or rather into an II-link through infection. It follows that
\(\langle k_S\rangle=\langle k_I\rangle\) in DE if and only if \(w=p\). Remarkably, for \(w=p\) not only the
average degrees of the two subensembles, but also the overall degree distributions of the S-nodes and the I-nodes coincide for MC simulations in DE, a feature that escapes the simplified description of Eqs. \ref{e:pm}. Hence when disease propagation along a SI-link and its rewiring are equally probable, a symmetry between the two node subensembles ensues.

The onset of the simple endemic phase, approximated in \cite{Gross:2006}, can be exactly given as
\[\frac{w}{w+p+r} <\frac{p \langle k \rangle -r}{p\left( \langle k\rangle+1\right)} ,
\]
confirming that rewiring increases the epidemic threshold. Furthermore it can be shown that, for any given \(\langle k \rangle > 1\), \(w> p+r\) is a necessary condition for the advent of bistability or an oscillatory regime. Consequently a qualitive departure from classic SIS dynamics without rewiring is only possible in the pairwise framework if topology change outweighs disease dynamics. 

While Eqs. \ref{e:pm} give a coarse-grained description of the adaptive network through various averages, the joint steady-state subensemble degree distributions \(P_{S,I}(x,y)\) of S- and I-nodes provide a more detailed probabilistic account and are therefore aimed at by recent models \cite{Marceau:2010,us:2011}. Since one can extract various network motif densities from the \(P_{S,I}(x,y)\), these distributions also encode average lifetimes of any node and link type in a network in DE (as do Eqs. \ref{e:pm} by generating triplet densities through the moment closure): In DE, a node is on average susceptible for \(\tau_S=[S]/\left(p[SI]\right)\), while the mean lifetime \(\tau_{SI}\) of an SI-link is determined by considering all of its decay channels, i.e. recovery of the infected node, breakup through rewiring and infection of the susceptible partner. Hence
\[\frac{1}{\tau_{SI}}=r+w+p\left(2\frac{[ISI]}{[SI]}+1\right).\]
Similarly, one obtains \(\tau_{SS}=[SS]/\left(p[SSI]\right)\) and \(\tau_{II}=1/\left(2r\right)\) for the mean lifetimes of SS- and II-links, respectively, with all motif densities taken from the network in DE.

\section{Node Cycle}
\label{s:nc}
 
A new \emph{stochastic} formalism was proposed in \cite{us:2011} to analytically describe a wide class of 
adaptive networks in DE. Under the assumption that the topology-changing mechanism ensures ergodicity in node-state and node-degree evolution, the \emph{node cycle} (NC) extracts node-ensemble statistics from a single node's long-term behavior in DE, treating its joint degree evolution in each state as a random walk. These random walks are coupled by the state-change 
rules and guided by the set \(\mu\) of model parameters defining local node dynamics, as well as by the 
set \(\kappa\) of \emph{correspondence parameters} that approximate the influence of the network 
background on local node dynamics in DE. For each \(\mu\) and \(\kappa\), an analytic expression in 
closed form can be obtained for the transition matrix of the Markov chain in joint degree space 
associated with a full cycle along the consecutive node states. From its Perron-Frobenius eigenvector, whose existence 
and uniqueness is model-intrinsic, a range of probability distributions can be derived describing the 
stationary state and degree probability distributions for \emph{arbitrary} values of the correspondence 
parameters. These are for each node state \(A\) the node's distribution of initial joint degrees \(\Phi^*_A\), its joint degree distribution \(P^*_A\) and its stage lifetime distribution \(T^*_A\). Both \(P^*_A\) and \(T^*_A\) are linear tranformations of \(\Phi^*_A\) in the respective node stage \(A\) \cite{us:2011}.

The correspondence parameters must be chosen so as to fulfill any desired global properties on the 
network the NC is unable to account for, such as for instance a fixed global average degree. 
Embedding local node dynamics into a consistent network background imposes additional independent 
constraints on the correspondence parameters. Fulfilling these constraints by an appropriate choice of \(\kappa\) poses a well-defined, albeit convoluted optimization problem (see Sec. \ref{s:corr}). With the optimal choice for \(\kappa\) the NC becomes a fully self-contained analytical framework for which the long-term dynamics of a single node is made consistent with a given network background.
Once this has been ensured, the aforementioned distributions represent the network's node-ensemble statistics. Important averages can then be obtained from them to complement an extensive probabilistic description of the adaptive network in DE: Considering lifetime and degree distributions delivers steady-state densities of low-order network motifs, as well as mean node-state and link lifetimes \cite{us:2011}.

\section{SIS with Selective Rewiring in the Node Cycle}
\label{s:me}

For the SR mechanism, the Master equations of the joint degree evolution of a single node in a network in DE in stage \(A\) (\(A \in \{S,I\}\)) can be set up. With \([x,y]\equiv P_A(x,y,t|x_0,y_0)\) being the probability that, having started off with \(x_0\) susceptible and \(y_0\) infected neigbors, the node under consideration possesses a joint degree of \(x\) susceptible and \(y\) infected neighbors at time \(t\), the node's susceptible \emph{S-stage} reads as \cite{us:2011}
\begin{align}\label{e:meS}
\frac{d[x,y]}{d t}  =& \{w+r\}\{\left(y+1\right)[x-1,y+1]-y[x,y]\} \nonumber\\ 
& - p\ y[x,y] +\tilde{w}\left([x-1,y]-[x,y]\right)\nonumber \\  
& +\tilde{p}_S\{\left(x+1\right)[x+1,y-1]-x[x,y]\},
\end{align}
and in the infected \emph{I-stage} as
\begin{align}\label{e:meI}
\frac{d[x,y]}{d t}  =& r\{\left(y+1\right)[x-1,y+1]-y[x,y]\} \nonumber\\ 
&+ w\{\left(x+1\right)[x+1,y]-x[x,y]\}-r[x,y] \nonumber\\  
& +\tilde{p}_I\{\left(x+1\right)[x+1,y-1]-x[x,y]\}.
\end{align}
Here the boundary conditions are \([-1,y]=[x,-1]=0\) in both the S- and I-stage, and the sets of model and correspondence parameters are \(\mu=\{w,p,r\}\) and \(\kappa=\{\tilde{w},\tilde{p_S},\tilde{p_I}\}\), respectively. The correspondence parameters \(\tilde{p}_{S,I}\) approximate the force of infection on susceptible neighbors in the two stages, and \(\tilde{w}\) the net degree gain through being-rewired-to. The state and degree evolution of a single node in a network in DE is then given by a composite random walk described with Eq. \ref{e:meS} in the S- and Eq. \ref{e:meI} in the I-stage. For \(\tilde{p}_{S}=\tilde{w}=0\), i.e. with frozen network dynamics as described by the NC, the absorbing state of that random walk is any \([x,0]\) (\(x \in \mathbb{N}\)) in the S-stage and corresponds to the system's disease-free equilibrium also described by Eqs. \ref{e:pm}.

For each \(\mu\) and \(\kappa\), the steady-state probability distributions \(P^*_{S,I}(x,y)\), \(\Phi^*_{S,I}(x,y)\) and \(T^*_{S,I}(t)\) of a node's degree evolution are calculated from Eqs. \ref{e:meS} and \ref{e:meI}. Setting a cutoff for the maximum total degree \(k_\text{max}\) considered in them, this computation is reduced to an eigenvector problem \cite{us:2011}. Distributions \(P^*_{S,I}(x,y)\) yield the mean number \(\langle B \rangle_A\) of adjacent B-nodes in the A-stage, and similarly one obtains \(\langle BC \rangle_A\), the mean number of (open or closed) triplets with the central node in stage \(A\) connected to a B- and C-node (\(A,B,C \in \{S,I\}\)). The first moments of \(P^*_{S,I}(x+y)\) are consequently given by \(\langle k^*_{S,I}\rangle=\langle S \rangle_{S,I}+\langle I \rangle_{S,I}\) and are the mean degrees of the node in the respective stage. The survival functions \(L^*_{S,I}(t)\) of the two node subensembles are defined as
\[L^*_{S,I}(t)=1-\int\limits_{0}^{t}T^*_{S,I}(t')\, \mathrm{d}t',\]
with the mean duration \(\tau^*_{S,I}\) of either stage obtained through
\[\tau^*_{S,I}=\int\limits_0^{\infty} t\, T^*_{S,I}(t) \, \mathrm{d} t,\] 
which for the I-stage yields \(\tau^*_I=1/r\) as expected \cite{us:2011}.

Choosing \(\kappa\) so that the actual network process in DE is mirrored by Eqs. \ref{e:meS} and \ref{e:meI}, distributions \(\Phi^*_{S,I}\), \(P^*_{S,I}\), \(T^*_{S,I}\) and \(L^*_{S,I}\) describing a node's degree evolution become distributions \(\Phi_{S,I}\), \(P_{S,I}\), \(T_{S,I}\) and \(L_{S,I}\) also encapsulating node-ensemble behavior in the respective DE. Similarly, NC averages \(\tau^*_{SI}\) and \(\langle k^*_{SI}\rangle\) then accurately represent mean subensemble lifetimes \(\tau_{S,I}\) and mean degrees \(\langle k_{SI}\rangle\). In DE, the \emph{prevalence} \([I]\) must be equal to the fraction of the time a node typically is infected, and therefore
\begin{equation}\label{e:i}
\tau_S/\tau_I=\left(1-[I]\right)/[I]
\end{equation}
must hold, allowing to obtain node-state densities from the NC.
\section{Optimal Correspondence Parameters and Network Stationary States}
\label{s:corr}
To tie equilibrium dynamics as described by the NC framework to the actual network process, several constraints on the choice of correspondence parameters \(\kappa\) need to be imposed. The \(\mu\) and \(\kappa\) describing a particular DE can then be used to generate characteristic probability distributions within the NC and compare them to the output of individual-based simulations (see Sec. \ref{s:compR}). Demanding for link conservation (a constant mean degree \(\langle k \rangle\)) yields \(\langle k \rangle=\left(\tau_S \langle k_S^*\rangle+\tau_I \langle k_I^*\rangle\right)/\left(\tau_S+\tau_I\right)\) and thus
\begin{equation}\label{e:ck}
C_0\equiv \left(1-\frac{\left(\tau_S \langle k_S^*\rangle+\tau_I \langle k_I^*\rangle\right)/\left(\tau_S+\tau_I\right)}{\langle k\rangle}\right)^2=0.
\end{equation}
For a choice of \(\kappa\) that ensures correspondence between NC and network dynamics in DE, the three correspondence parameters introduced in Sec. \ref{s:me} are similar in nature, in that they describe the number of infected neighbors of S-nodes, either without additional conditions or given that the latter are attached to another particular node type. While then \(\tilde{p}_{S,I}/p\) approximates the mean number of infected neighbors of a susceptible neighbor in the respective stage, \(\tilde{w}/w\) yields the mean number of infected neighbors of a S-node regardless of the neighborhood of the latter. Hence, in the network description, each of these terms can be expressed by a mean field composed of low-order motif densities. At exact correspondence, those densities are also given by the NC averages introduced in Sec \ref{s:me}. Consequently, the following self-consistency relations should hold regardless of imposed mean degree: 
\begin{align}\label{e:ctilde}
C_1\equiv & \left(1-\frac{\langle I \rangle_S}{\tilde{w}/w}\right)^2 =0 \nonumber\\
C_2\equiv &\left(1-\frac{\langle SI \rangle_S/\langle S\rangle_S}{\tilde{p}_S/p}\right)^2 =0 \nonumber\\
C_3\equiv &\left(1-\frac{2\langle II \rangle_S/\langle I\rangle_S+1}{\tilde{p}_I/p}\right)^2 =0.
\end{align}
The NC averages in Eqs. \ref{e:ck} and \ref{e:ctilde} are convoluted functions of \(\mu\) and \(\kappa\) evaluated within the NC, and thus for a given adaptive network with fixed \(\mu\) and \(\langle k \rangle\), each \(C_i\) is a cost function in \(\kappa\) defined by the respective constraint. Because deciding for constant \(\tilde{p}_{S,I}\) only approximates the interaction of a node's neighborhood with the rest of the network, these four constraints may not admit for a solution in \(\kappa\) at all, not even for any single constraint. Hence establishing correspondence with network dynamics in DE is achieved through identifying \emph{optimal} \(\kappa\) that minimize all \(C_i\). 

\begin{figure}[ht]
\centering
  \includegraphics[width=\textwidth]{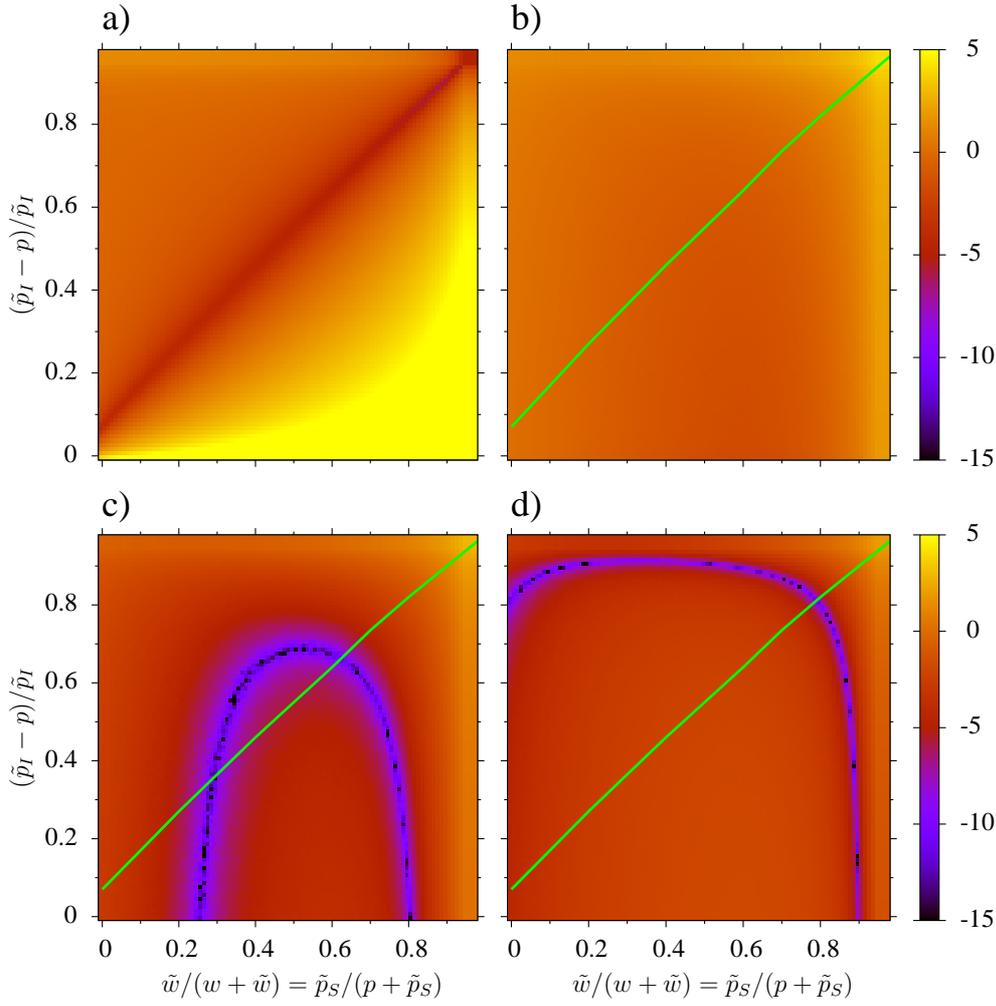}
   \caption{(Color online) Logarithmic color-coded plots of cost functions \(C_i\) on \(\tilde{w}/w=\tilde{p}_S/p\) plane with axes rescaled to unit length. Coordinates for low cost function values (\(\leq 10^{-5}\)) are \(\kappa\) ensuring good correspondence for the respective constraint. a): \(C_1+C_2+C_3\), with optimizing \(\kappa\) close to, but not on \(\tilde{w}/w=\tilde{p}_S/p\). b)-d): \(C_0\), with solid lines representing \(\kappa\) that minimize \(C_1+C_2+C_3\) as in a). b): \(\langle k \rangle=3\). c): \(\langle k \rangle=5\). d): \(\langle k \rangle=7\). Model parameters \(w=0.05\), \(p=0.008\), \(r=0.005\) and maximum cutoff degree \(k_\text{max}=50\).}
\label{f:cf}
\end{figure}

Due to the different nature of Eq. \ref{e:ck} on the one and Eqs. \ref{e:ctilde} on the other hand, it is instructive to first compute the set of \(\kappa\) minimizing \(C_0\) and of those \(\kappa\) minimizing \(C_1+C_2+C_3\) separately. In the absence of degree-status correlations and a characteristic degree imposed on the system by minimizing \(C_0\), each of the \(C_i\), \(i=1,2,3\), would be minimal along  
 \(\tilde{w}/w=\tilde{p}_{S}/p=\tilde{p}_{I}/p-1\). Indeed, MC simulations confirm that the roots of \(C_1\) and \(C_2\) for the motif densities computed from the network in DE verify
 \(\tilde{w}/w\approx\tilde{p}_{S}/p\) in a wide range of \(\mu\). Thus, for visualization purposes we shall represent the cost functions on the  \(\tilde{w}/w=\tilde{p}_{S}/p\) plane of parameter space (Fig. \ref{f:cf}).

It can be seen in Fig. \ref{f:cf}a) that, the lower \(w\) is chosen to weaken degree-status correlations, the closer the optimal \(\kappa\) for \(C_1 + C_2 + C_3\) is to the diagonal \(\tilde{w}/w=\tilde{p}_{S}/p=\tilde{p}_{I}/p-1\). Since candidates for globally optimal correspondence parameters should ensure minimization of all cost functions, one has to identify regions of overlap of those \(\kappa\) which minimize either only \(C_0\) or \(C_1\) to \(C_3\). Fixing \(\mu\) and changing the imposed mean degree via Eq. \ref{e:ck} then allows for browsing the system's phases as defined by different numbers of overlap regions (Figs. \ref{f:cf}b)-d)). No region of overlap indicates 
that the NC cannot be  matched to a network in DE, and corresponds to parameter values for which the network is in the 
frozen phase (Fig. \ref{f:cf}b), low \(\langle k \rangle\)).
One overlap region corresponds to a simple endemic phase (Fig. \ref{f:cf}d), high \(\langle k \rangle\)). Two regions of overlap are associated with a bistable phase (Fig. \ref{f:cf}c), intermediate \(\langle k \rangle\)): The \(\kappa\) with larger components represents the stable DE of the stable active branch, whereas the one with smaller is associated with the \emph{unstable} DE of the hysteresis loop leading to bistability in Eqs. \ref{e:pm} \cite{Gross:2006}.

The approximate values for the \(\kappa\) minimizing \(C_0\) to \(C_3\) (i.e the coordinates of overlap regions in the \(\tilde{w}/w=\tilde{p}_{S}/p\) plane) are used as initial guesses for the minimization of the overall cost function \(C_0+C_1 + C_2 + C_3\) in full correspondence parameter space. Standard minimization routines then yield the optimal correspondence parameters used in the following section to obtain the NC results.

Since Eqs. \ref{e:ctilde} also allow for a reverse calculation of optimal \(\kappa\) for given set of motif densities of a network in DE, they can be used to approximate the expected coordinates of optimal \(\kappa\) in every overlap region. For stable DEs of a variety of \(\mu\) and imposed \(\langle k\rangle\), these predicted coordinates very well match those obtained within the NC. Using the moment closure approximation in \cite{Gross:2006}, steady-state densities of triplet motifs can be calculated in the pairwise model alongside node-state and link densities. For unstable DEs in the bistable phase of Eqs. \ref{e:pm}, these densities again give a good estimate for the coordinates of the respective overlap region, underlining the validity of the NC approach.

Given the same \(\mu\) in the NC and MC simulations, the values of imposed mean degree \(\langle k \rangle\) that trigger a change in the number of overlap regions coincide with the values of \(\langle k \rangle\) observed at corresponding phase transitions in simulations. Furthermore this number is equal to the number of stable and unstable DEs predicted by Eqs. \ref{e:pm} for the corresponding phase. Hence the NC allows for the localization of all stable \emph{and} unstable dynamic equilibria for a given SIS adaptive network. It is moreover able to structurally explore the dynamics by correctly predicting the existence of and the transitions between phases with different numbers of DEs. The supercritical Hopf bifurcation in Eqs. \ref{e:pm} that gives rise to a small stable oscillatory regime in the bistable phase \cite{Gross:2006} leaves the number of DEs unchanged. Consequently no indications for that phase transition were found in the NC, as the framework cannot distinguish between the bistable phase's active steady state and its oscillatory regime. Concurrently, the NC is insensitive to the stability of its detected DEs. Yet it strongly hints that unstable DEs are physical and not mere model artefacts, and offers, unlike current compartmental models 
and MC simulations, an extensive description of them.

\section{Comparison of Rewiring Mechanisms}
\label{s:compR}
Two modifications of the adaptive SIS model are presented here to showcase the applicability of the NC framework. The first, \emph{media-driven rewiring} (MR) introduced in \cite{Gross:2008}, relates disease awareness to instantaneous knowledge of the prevalence \(\tilde{i}\) through a rewiring rate \(w \cdot{\tilde i}\) (\(w=\text{constant}\)). Hence in MR, a global time-dependent quantity feeds back to the rate of a semi-local rewiring mechanism. The second modification, proposed in \cite{Zanette:2008}, suggests that susceptibles rewire links with a constant rate to a randomly selected node regardless of the state of the latter. This \emph{blind rewiring} (BR) serves as an antipode to the strictly selective rewiring put forward in the original model presented in Sec. \ref{s:sis}, with parametrizations interpolating between those two limiting cases acknowledging partial knowledge of other individuals' disease status in a population.

To model MR in the NC, the additional correspondence parameter \(\tilde{i}\) has to be introduced, and all rewiring terms in Eqs. \ref{e:meS} and \ref{e:meI} must be rescaled by a factor \(\tilde{i}\). This new correspondence parameter is also needed to properly describe an additional mean field necessary for BR. Since there a SI-link is rewired to an I-node with probability \(\tilde{i}\), both rewiring terms in Eq. \ref{e:meS} for the S-stage need to be rescaled by a factor \(\left(1-\tilde{i}\right)\) to represent "successful" rewiring to another S-node. In Eq. \ref{e:meI} for the I-phase no such rescaling takes place, but its existing rewiring term ought to be complemented by adding \(\tilde{w}\left(1-\tilde{i}\right)\{[x-1,y]-[x,y]\}\) to account for the "erroneous" linking of S- to I-nodes. With Eqs. \ref{e:ck} and \ref{e:ctilde} still being a valid set of constraints for both MR and BR, \(\tilde{i}=[I]\) should hold, resulting in the additional constraint
\begin{equation}\label{e:ci}
C_4\equiv\left(1-\frac{\tau^*_I/\left(\tau_S^*+\tau^*_I\right)}{\tilde{i}}\right)^2=0
\end{equation}
through Eq. \ref{e:i}. Again the mean lifetimes in Eq. \ref{e:ci} are convoluted functions of model parameters \(\mu=\{w,p,r\}\) and correspondence parameters \(\kappa=\{\tilde{w},\tilde{p_S},\tilde{p_I},\tilde{i}\}\) evaluated entirely within the NC. Therefore in the two modified rewiring scenarios, cost functions \(C_0\) to \(C_4\) ought to be minimized to establish correspondence between the NC and the respective network model in DE. A simple calculation shows however that in both MR and BR
\[\tilde{i}=\frac{1}{wr/\left(\tilde{w}p\right)+1}\]
holds for every optimal \(\kappa\), so that \(\tilde{i}\) is completely determined by the other correspondence and model parameters. Therefore the search space for any optimization algorithm aimed at fulfilling constraints \(C_0\) to \(C_4\) remains three-dimensional.

\begin{figure}[ht]
\centering
  \includegraphics[width=\textwidth]{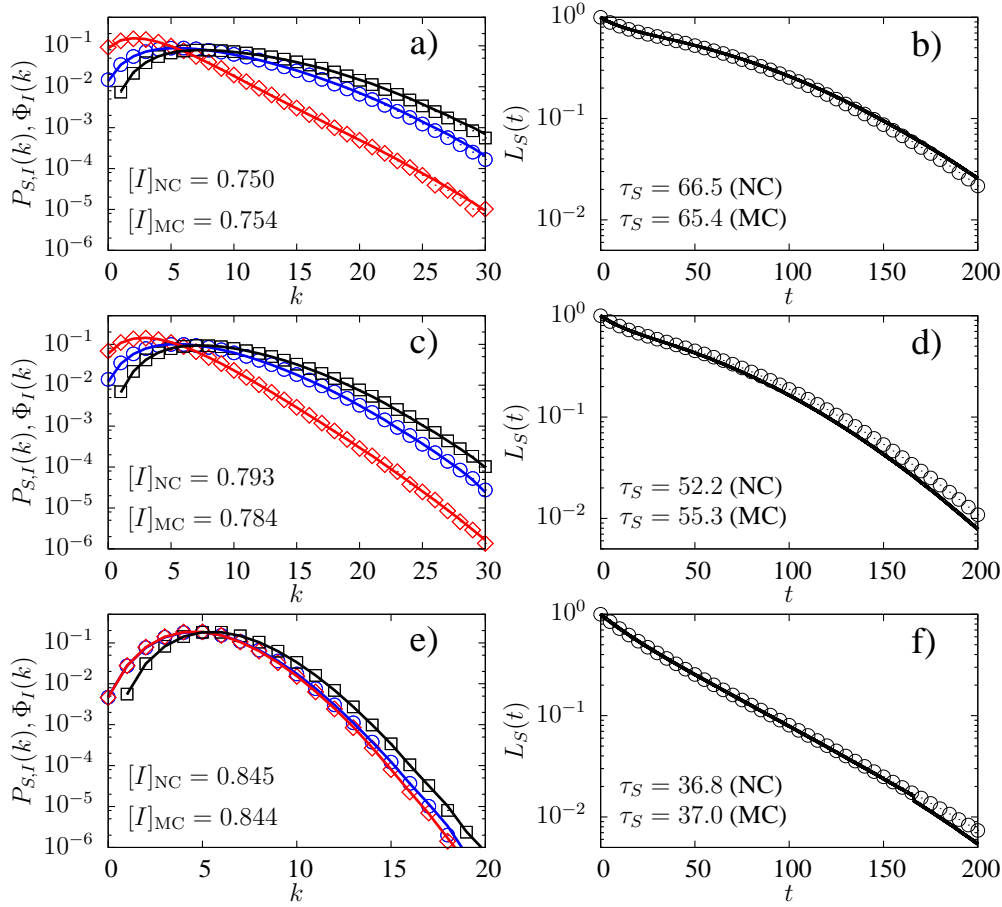}
  \caption{
(Color online) Characteristic distributions of a network in DE for SR (a)-b)), MR (c)-d)) and BR (e)-f)). Left column: Degree distributions for S-nodes (circles), I-nodes (diamonds), as well as of initial degrees of I-nodes (squares). Right column: Plots of survival functions of S-nodes. Solid lines are predictions by the NC. Insets: Comparison of prevalence \([I]_\text{MC}\) taken from MC simulations and \([I]_\text{NC}\) computed by in the NC (left column), and of mean S-lifetimes obtained from MC simulations and the NC (right column).
\(w=0.05\), \(p=0.008\), \(r=0.005\); \(\langle k \rangle=5\),  \(k_\text{max}=80\). 
a)-b): SR (stable active branch of bistable phase) with \(\tilde{w}=0.095\), \(\tilde{p}_S=0.017\), \(\tilde{p}_I=0.027\).
c)-d): MR (simple endemic phase) with \(\tilde{w}=0.12\), \(\tilde{p}_S=0.022\), \(\tilde{p}_I=0.031\). 
e)-f): BR (stable active branch bistable phase) with \(\tilde{w}=0.17\), \(\tilde{p}_S=0.026\), \(\tilde{p}_I=0.035\). 
MC simulations according to \cite{Gillespie:1976} with \(N=5\cdot 10^4\) nodes, initial Erd\H{o}s-R\'{e}nyi graph and initial prevalence \([I]_0=0.6\). Statistics were recorded at \(t=2\cdot 10^4\) for \(10^4\) network realizations. Error bars are smaller than markers.
}\label{f:DD}
\end{figure}

Having been provided with the set of model parameters and the mean degree specifying an adaptive network, the NC identifies the DEs and the corresponding sets of optimal correspondence parameters. For all three considered rewiring mechanisms, one can then extract the various probability distributions and averages described in Secs. \ref{s:nc} and \ref{s:me}, and compare them to the output of 
MC simulations (Fig. \ref{f:DD}). Due to recovery being neighborhood-independent and happening at a constant rate, \(P_I(x,y)=\Phi_S(x,y)\) in all three scenarios, i.e the steady-state degree distribution of I-nodes is identical to the distribution of their final degrees right before recovery \cite{us:2011}.

For the model parameters and mean degree used in Fig. \ref{f:DD}, both SR and BR are in the bistable phase \cite{Gross:2006,Zanette:2008}. The respective DE is chosen to be in the stable active branch, whereas for the same choice of parameters, MR is in its simple endemic phase \cite{Gross:2008}. There, MR is equivalent to SR with the rewiring rate rescaled by a factor \([I]\) (the network's steady-state prevalence). Consequently degree distributions and survival functions of those two rewiring mechanisms resemble each other (Figs. \ref{f:DD}a)-d)), with the higher effective rewiring rate in SR prolonging the S-stage, lowering overall prevalence and letting both subensembles sample large degrees in comparison with MR. The higher the overall steady-state prevalence, the more similar are those two rewiring scenarios.

At very low prevalences however, the SR and BR model essentially describe the same behavior, since they both feature approximately the same rate of successful rewiring to another S-node. If chosen model parameters and mean degree force SR into the bistable phase, there will be no low-prevalence active steady state in SR \cite{Gross:2006}, and thus neither in BR. Hence for the same set of \(\mu\) and \(\langle k \rangle\), the prevalence in a steady-state BR has to be high, and consequently successful rewiring happens on a much slower timescale than disease dynamics governed by model parameters \(p\) and \(r\). Topological separation of node states (through rewiring) is consequently outpaced by their randomization (through conventional SIS dynamics), so that rewiring effectively randomizes network topology. Much like for very low rewiring rates in SR, Poissonian degree distributions around the overall mean degree ensue (Fig. \ref{f:DD}e)), with almost no node-status clustering occuring (as opposed to SR and MR with same \(\mu\) and \(\langle k \rangle\)).

The exponential survival function in Fig. \ref{f:DD}f) implies exponential lifetime distributions \(T_S(t)\) of S-nodes for BR, indicating that the force of infection on a S-node, and thus its number of infected neighbors, is approximately constant in DE.  Of the three processes that influence that number, both recovery and successful rewiring have very low rates, the latter due to the large prevalence \([I]\). The rate of the third process - conversion of
susceptible into infected neighbors - is also small due to the small number 
of susceptible neighbors of a typical node:  Neither in its S-stage (due to erroneous rewiring and slow recovery of infected neighbors) nor in the preceding I-stage (due to again a modest recovery rate and a large number of competing I-nodes at the receiving end of erroneous rewiring) can the node agglomorate a large fraction of susceptible neighbors. Consequently the relative change in the number of infected neighbors is small, and hence the neighborhood of a S-node exerts an almost constant force of infection on it.

Since the iterative nature of the optimization procedure outlined in Sec. \ref{s:corr} makes the latter computionally expensive (see also Sec. \ref{s:compF}), a relatively small cutoff of total degree \(k_\text{max}\) in NC transition matrices is convenient to identify DEs. However, properly modeling the degree evolution of a node with Eqs. \ref{e:meS} and \ref{e:meI} requires setting a sufficiently large \(k_\text{max}\). Consequently optimal \(\kappa\) obtained for low degree cutoffs may provide a slightly blurred correspondence between the NC framework and an adaptive network in DE, resulting in small deviations in distributions and averages as observed in Fig. \ref{f:DD}.

\section{NC Computation and Comparison to other Frameworks}
\label{s:compF}
In the case of large cutoff degrees \(k_\text{max}\) or a high number \(n\) of node states, the procedures involved in the NC, as laid out in Secs. \ref{s:nc} to \ref{s:corr} and \cite{us:2011}, can be computationally expensive: For each of the \(n\) node stages, the Master equation for the respective joint degree evolution is set up and solved for arbitrary \(\mu\) and \(\kappa\). For every \(\kappa\) considered, one subsequently computes \(n\) dense stage-transition matrices with \(\sim k_\text{max}^{2n}\) nontrivial coefficients each. Then the unique positive eigenvector of the resulting matrix product yields a probability distribution (\(\Phi_S^*(x,y)\) in the SR scenario) and consequently also steady-state network motif densities needed for the cost functions' evaluation at \(\kappa\), allowing for the functions' minimization and ultimately identifying optimal correspondence parameters of the DE. A more straightforward route to those essential network motif densities is to write down the Master equation for the composite random walk through all node stages, i.e. the transition matrix for the full node cycle. The resulting equations are equivalent to a stochastic interpretation of the compartmental formalism presented in \cite{Marceau:2010} if time-independent mean fields of link gain in the S-stage and of infection acting upon susceptible neighbors are assumed. This transition matrix has \(\sim\left(n k_\text{max}^{n}\right)^2\) entries, but is sparse, and its nonzero entries are linear combinations of model and correspondence parameters. By arguments similar to those brought forward in \cite{us:2011}, the matrix' null space is one-dimensional and yields steady-state degree distributions as well as node-state densities (in SR \(P_{S,I}^*(x,y)\) and \([I]\), respectively), directly delivering the crucial network densities needed in the quest for optimal \(\kappa\). The computational effort involved in evaluating the cost functions in this modified NC thus can be significantly reduced compared to implementing the classic NC procedure.
\begin{table}[h]
\begin{tabular}{ p{0.19\textwidth}  | p{0.22\textwidth}  | p{0.24\textwidth} | p{0.22\textwidth} }
 & \textbf{Pairwise Model} & \textbf{Compartmental Model} & \textbf{Node Cycle}\\
\hline
\emph{transient modeling} & numerical &  numerical  & - \\
\hline
\emph{DE detection} & analytical & (numerical) & (analytical)  \\
\hline
\emph{DE description} &  \([A]\),\([AB]\) & \([A]\), \(P_{A}\) & \([A]\), \(P_{A}\), \(\Phi_{A}\), \(T_{A}\) \\
\hline
\emph{DE stability} & stable/unstable DEs detected, described, distinguished & stable DEs detected,\(\quad\) described &stable/unstable DEs detected, described \\ 
 \end{tabular}
\caption{Comparison of frameworks capturing DEs in adaptive networks. Node-state densities \([A]\) and subensemble joint degree distributions \(P_A\) encode link densities \([AB]\), higher-order star motif densities as well as mean node-state and link lifetimes (see Sec. \ref{s:sis} and \cite{us:2011}), with \(A\) and \(B\) assuming any value in node-state space.  For every node stage \(A\), \(T^*_A\) and \(\Phi^*_A\) are the stage lifetime distribution and joint distribution of initial degrees, respectively. Identifying DEs with the compartmental model is subject to the usual limitations of numerical integration, whereas for the NC, cost functions are computed analytically and minimized with
standard optimization techniques.}
\label{t:comp}
\end{table}
The three main modeling frameworks dealing with dynamic equilibria in adaptive networks are contrasted in Tab. \ref{t:comp}. To account for correlations beyond immediate network neighbors, the pairwise model from \cite{Gross:2006} can in principle be extended to a moment closure at the level of higher-order network motifs, at the expense of analytical tractability due to an increasing number of nonlinear ODEs involved. Similarly, the accuracy of the compartment model in \cite{Marceau:2010} and the NC can be improved, with computation times drastically increasing.

By construction, the pairwise model is too coarse-grained to distinguish between some vital differences in rewiring mechanisms, whereas more local frameworks like the NC or the compartmental model can accommodate those changes that do alter the ensuing steady-state topology. If for instance a SI-link has been cut in the original SR scenario, adding it between \emph{two} randomly selected susceptibles instead of classically rewiring it to just one implies a different random walk in the NC than Eqs. \ref{e:meS} and \ref{e:meI}, and hence altered degree distributions extracted from them. The equations of the pairwise model however cannot account for this change in rewiring mechanism and remain equal to those describing the SR scenario. Likewise, BR reduces like MR to a rescaled SR in Eqs. \ref{e:pm}, whereas in the node cycle and the compartmental model, that structural difference between BR and SR is correctly accounted for.
Moreover the NC in its current form is, unlike the two ODE-based models, not designed to capture active phase dynamics other than DEs. But similar to the pairwise model, it can indicate a global frozen phase for given model parameters \(\mu\): When the NC's different cost functions do not display overlapping minima, no self-consistent embedding of single-node dynamics in a network in DE exists (see. Fig. \ref{f:cf}b)).
\section{Conclusion}
\label{s:con}
In this paper, we review different frameworks dealing with cyclic node-state dynamics on adaptive networks. Focusing on the SIS model with rewiring, we discuss both previous and new results that can be derived in the framework of its low dimensional deterministic description through the
standard pair approximation. By its very nature, this approach cannot describe one of the most striking
features of the model. This is the fact that for a large parameter regime, simulations on networks show the
states of the nodes and the links coevolving to produce and maintain a dynamic
network topology characterized by well-defined degree distributions not only
for the global network, but also for the subsets of S- and I-nodes.

The main focus of the paper is describing these steady-state network statistics, for they are essential to understanding
the relation between dynamics and network structure and to apply these ideas to
real world examples. We briefly present
the ideas of a method that has been proposed recently to obtain an analytic
description of the dynamic equilibria of adaptive networks. We then apply this method, called the node
cycle, to the SIS model with three different types of rewiring. These
three different rewiring schemes cover a whole range of assumptions about the
less well-established "social" dynamics of the model.

A key point of the node cycle method is the determination of the
"correspondence parameters", whose role is to embed the general node-cycle
description into the dynamics of a particular network through the fulfilment of
self-consistency conditions that express the network's topology in dynamic equilibrium. We discuss in detail how to determine the optimal values for these
correspondence parameters within the node cycle framework. For a given set of model parameters, this self-consistency approach detects and describes all dynamic equilibria predicted by the pairwise model, notably including a second equilibrium in the bistable phase that can be identified with an unstable active steady state of the system.

The node cycle method allows the determination not only of the degree
distributions for all node stages, but also of their distributions in initial degrees, their lifetime profiles and of the
densities of low-order network motifs in the steady state. We contrast its scope of description of dynamic equilibria with that of both the compartmental and the pairwise model and identify \emph{areas of competence} for each of the descriptions. The computational effort involved in the node nycle model is laid out, and an alternative, faster route to aforementioned steady-state characteristics is given. For the three rewiring schemes that we considered, the results for the steady-state degree distributions and lifetime profiles derived from the node-cycle method are in very good agreement with the results of simulations on networks. As to the comparison of the different rewiring schemes, we find that
modifications of the standard rewiring rules with very similar functional forms
can have either negligible or dramatic effects on the resulting network topology. 

In conclusion, we review several frameworks dealing with dynamic equilibria in adaptive networks, elaborate on the pairwise formalism of Gross et al. that models SIS dynamics on adaptive networks, apply the node-cycle method and show that it can be used to identify steady states. We subsequently describe the latter through degree distributions and
other steady-state statistics. In future work we intend to explore the method further by applying it to other processes on adaptive networks beyond a binary state space.
\subsection*{ }
Financial support from the Portuguese Foundation for Science and Technology (FCT) under Contract POCTI/ISFL/2/261 is gratefully acknowledged. The first author was also supported by FCT under Grant No. SFRH/BD/45179/2008. A.P. additionally thanks FCT for financial support through the Ci\^encia 2007 program.

\bibliographystyle{unsrt}
\bibliography{wielandref}

\end{document}